\documentclass[twocolumn,showpacs,preprintnumbers,prb,amsmath,amssymb]{revtex4}
\usepackage{graphicx}
\usepackage{dcolumn}
\usepackage{bm}
\usepackage{longtable}
\usepackage{array}
\usepackage{color}

\begin{document}

\title{Spectroscopic signatures of many-body interactions and delocalized states in self-assembled lateral quantum dot molecules}
\author{X. Zhou$^{1}$}
\author{S. Sanwlani$^{1}$}
\author{W. Liu$^{1}$}
\author{J. H. Lee$^{2,3}$}
\author{Zh. M. Wang$^{3,4}$}
\author{G. Salamo$^{3}$}
\author{M. F. Doty$^{1}$}
\email{doty@udel.edu}

\affiliation{$^{1}$Dept. of Materials Science and Engineering, University of Delaware, Newark, DE 19716, USA}
\affiliation{$^{2}$School of Electronics and Information, Kwangwoon University, Nowon-gu, Seoul 139-701, South Korea}
\affiliation{$^{3}$Institute of Nanoscale Science and Engineering, University of Arkansas, Fayetteville, Arkansas 72701, USA}
\affiliation{$^{4}$State Key Laboratory of Electronic Thin Films and Integrated Devices, University of Electronic Science and Technology of China, Chengdu 610054, People's Republic of China}

\date{\today}

\begin{abstract}
Lateral quantum dot molecules consist of at least two closely-spaced InGaAs quantum dots arranged such that the axis connecting the quantum dots is perpendicular to the growth direction. These quantum dot complexes are called molecules because the small spacing between the quantum dots is expected to lead to the formation of molecular-like delocalized states. We present optical spectroscopy of ensembles and individual lateral quantum dot molecules as a function of electric fields applied along the growth direction. The results allow us to characterize the energy level structure of lateral quantum dot molecules and the spectral signatures of both charging and many-body interactions. We present experimental evidence for the existence of molecular-like delocalized states for electrons in the first excited energy shell.
\end{abstract}

\maketitle

Self-assembled semiconductor quantum dots (QDs) are a promising material for development of next-generation optoelectronic logic devices. A key advantage of these materials is the opportunity to confine single charges whose spin projections could serve as the logical bits for classical or quantum spin-based computing while preserving the opportunity to control these spin projections using ultrafast optical fields.  Quantum dot molecules (QDMs), composed of two coupled QDs, have unique properties that enable new methods of controlling and measuring spins.\cite{Doty2010, Liu2010, Vamivakas2010} Vertically-stacked quantum dot molecules (VQDMs), in which the two QDs are stacked along the growth direction, have received the most attention because the vertical growth protocol provides precise and independent control over the height of each QD and the intervening barrier.\cite{Bracker2006} In VQDMs, the formation of delocalized molecular states with symmetric and antisymmetric molecular orbitals has been clearly established.\cite{Krenner2005, Stinaff2006, Doty2006, Doty2009} The coherent tunneling that leads to the formation of these molecular states can also be used to create entanglement between two spins confined in the separate dots that compose a VQDM.\cite{Kim2010b} This approach, however, can not be extended to enable nearest-neighbor interactions for a large number of bit states. One obstacle is the inability to provide local gates between vertically-stacked QDs in order to independently control the interactions between an arbitrary pair of nearest neighbors. A second obstacle is that the growth-direction electric field that controls the coherent interactions cannot simultaneously be used to controllably charge each QD with a single electron or hole whose spin projection can serve as the logical bit.

Lateral quantum dot molecules (LQDMs) are an alternate QDM configuration in which closely-spaced QDs are formed on a single growth plane.\cite{Lee2006, Liang2006, Liang2008} The simplest LQDM is a diatomic-like pair of QDs that nucleate from a single site. LQDMs could be the basis of a scalable architecture based on coherent interactions between nearest-neighbors,\cite{Munoz2011} but the development and understanding of LQDMs lags behind VQDMs. One major obstacle is that the LQDM geometry makes it extremely challenging to deterministically create unique confining potentials for each QD. A second obstacle is that LQDMs typically have a relatively large center-to-center distance that significantly reduces the strength of tunnel coupling. Despite these challenges, LQDMs are an important testbed for the development of tools to deterministically electrically charge individual QDs in an array while preserving independent electrical and/or optical control over coherent nearest-neighbor interactions. As a first step toward the development of such tools, we embed InGaAs LQDMs in a three-terminal diode structure and investigate the spectroscopic signatures of charging and charge carrier interactions as controlled by the voltage applied to the diode in the growth direction. Our results provide experimental evidence that electrons in excited energy levels are delocalized over the entire LQDM.

In Sect.~\ref{Expt} we present the sample geometry and experimental approach. In Sect.~\ref{shells} we describe photoluminescence (PL) measurements of ensembles of LQDMs and propose a model for the LQDM energy shell structure in which electrons in excited energy shells are delocalized over the entire LQDM.  In Sect.~\ref{manybody} we describe the spectral shifts of the ensemble PL as the excited electron energy shell becomes occupied by additional electrons. Analysis of these results supports our energy shell model. In Sect.~\ref{single} we use spectroscopy of individual LQDMs to analyze the consequences of Coulomb interactions. The measured spectral shifts validate the energy shell model and suggest that the excited electron states are delocalized over the entire LQDM. In Sect.~\ref{finestructure} we report the observation of fine structure that likely originates in spin interactions within the excited electron energy shell. These observations further support the conclusion that the excited electron shell is delocalized over the entire LQDM. We conclude in Sect.~\ref{conclusion} by discussing how this energy shell structure creates new opportunities to control the interactions of spins confined in LQDMs.

\section{Experimental Method}\label{Expt}
\begin{figure}[htb]
\begin{center}
\includegraphics[width=8.0cm]{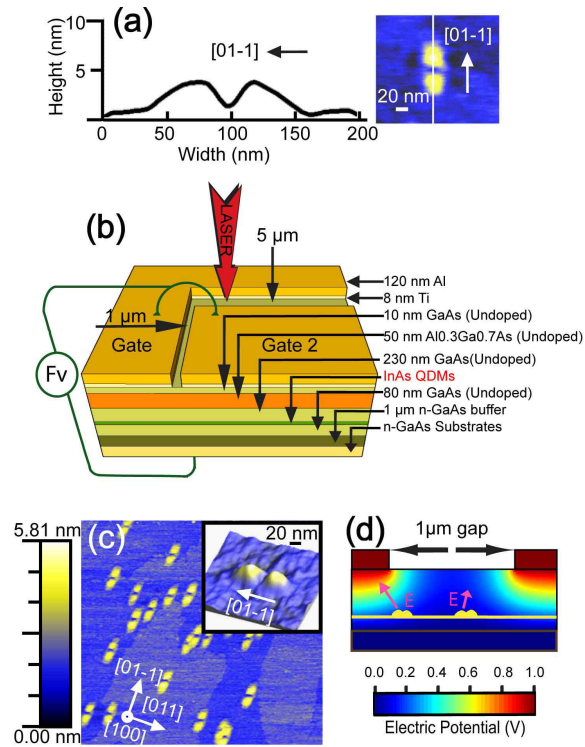}
\caption{(Color Online) a) Cross-sectional profile of a LQDM reveals the molecular geometry and height. The cross-section is along the axis shown in the inset. b) Sample and three-terminal electrode design. The Schottky diode structure permits application of electric fields along the growth direction. The interdigitated top contacts allow application of electric fields perpendicular to the growth direction and simultaneously isolates individual LQDMs for optical spectroscopy. c) AFM image of LQDMs grown under identical conditions to those embedded in the sample studied here. The large AFM image is 1$\times$1 $\mu$m area and the inset is the 3D AFM image in 200$\times$200 nm area. d) Potential profile of the three-terminal electrodes structure. The top two electrodes apply 1 V voltage while the bottom electrode applys 0 V. The magnitudes and directions of the electric fields along the two QDMs are marked with arrows respectively. \label{LQD}} \end{center}
\end{figure}

The InAs LQDMs we study are fabricated using solid source molecular beam epitaxy (MBE).\cite{Lee2010} First, InAs QDs are grown on GaAs using conventional Stranski-Krastanov self-assembly. The resulting dome-shaped QDs are partially capped with 10 ML of GaAs and then annealed \emph{in situ} for 30 s at $480^\circ$C. Anisotropic diffusion at the QD surface causes each single QD to evolve into a diatomic LQDM.\cite{Songmuang2003, Krause2005, Wang2008, Lee2010} The molecular axis is along the [0 1 -1] direction as shown in Fig.~\ref{LQD}a and c. A height profile of the LQDM, as shown in Fig.~\ref{LQD}a, indicates that the QDs are approximately 4 nm high. The LQDMs we study optically are capped with additional GaAs, which is expected to lead to some morphological changes in the LQDMs.\cite{Songmuang2003b}

The LQDMs studied here are grown in a diode structure in order to use applied electric fields to control the charge state of the molecules. The sample composition and layer thicknesses are shown in Fig.~\ref{LQD}b. The electric field is applied between an ohmic contact to the n-doped substrate and a Schottky contact to the top surface. The top contacts have an interdigitated pattern defined by photolithography and are composed of 120 nm of Al deposited on top of 8 nm of Ti by electron beam metal deposition. Because there are, on average, 30 LQDMs per $\mu$m$^2$, the $1 \mu$m narrow gaps in the interdigitated top contact serve as a mask that isolates individual LQDMs for optical spectroscopy. 5 $\mu$m gaps perpendicular to the 1 $\mu$m provide access for ensemble photoluminescence (PL) measurements. The interdigitated contacts provide the opportunity to apply both lateral (perpendicular to the growth direction) and vertical (parallel to the growth axis) electric fields in order to independently control the charge occupancy of the molecule and the quantum coupling between the two QDs that comprise the molecule. In this work we focus on charging of the QDMs and always apply the same voltage to both of the top contacts. Although the two interdigitated top contacts always have the same electrical potential, the long term goal of applying both lateral and vertical electric fields precludes the inclusion of a semitransparent Ti layer to maintain uniform electric fields across the  1 $\mu$m gap. The interdigitated contacts are separated by  1 $\mu$m, but only 370 nm separates the top contacts from the n-doped GaAs that provides the back contact. Consequently, the net electric field at the LQDM location contains both vertical and lateral components, as depicted in Fig.~\ref{LQD}d. The ratio of lateral and vertical field components depends on the spatial location of the LQDM within the lateral gap, which cannot be measured and varies from LQDM to LQDM. We therefore quote only the value of the applied voltage and not the net electric field.

For optical spectroscopy, the patterned sample was held in the cold finger of an Advanced Research Systems DMX-20 closed-cycle cryostat at a temperature of 12K. A Ti:Sapphire laser tuned to 870 nm was used for optical excitation. The PL signal was collected using a high numerical aperture objective, long-pass filtered to remove residual laser light, and analyzed with a 0.75 m spectrometer equipped with a liquid nitrogen cooled CCD camera. The spectral resolution of this system is 70 $\mu$eV.

\section{Energy shells of LQDMs}\label{shells}
Peng et al. have used atomistic empirical pseudopotential calculations of lens-shaped InGaAs LQDMs with a basin embedded in GaAs to calculate single particle energies of LQDMs.\cite{ Peng2010, Peng2010a} The approach used by Peng et al. models LQDMs grown by methods very similar to those used here. The calculations predict three groups of electron energy levels, with the ground ($E_0$) and first ($E_1$) excited energy shells separated by approximately 13 meV and the first ($E_1$) and second ($E_2$) excited energy shells separated by approximately 15 meV.\footnote{Note that Peng et al. use $E_0$, $E_1$, $E_2$, etc. to label each single particle energy state. We cannot assign all transitions to specific single particle or many body states. We use $E_0$, $E_1$, $E_2$ to refer to groups of energy levels analogous to the s, p, d shells of single QDs. The specific energy of each state, as predicted by Peng et al., varies with applied lateral electric field. We therefore quote only approximate values for the center of each energy shell.} The atomistic pseudopotential calculations predict that the $E_0$ electron states are isolated in the individual QDs, unless an applied electric field tunes the confined states of the two QDs into resonance. The pseudopotential calculations further predict that the $E_1$ and $E_2$ shells are delocalized over the entire LQDM for all lateral electric fields. In contrast, hole states within the first two energy shells ($H_0$, $H_1$) are predicted to be localized in a single QD over a wide range of lateral electric fields. Holes in the second excited energy shell ($H_2$) are predicted to show some degree of delocalization. The calculated ground ($H_0$) and first excited ($H_1$) states for holes are separated by about 15 meV, with the first ($H_1$) and second ($H_2$) shells separated by approximately 14 meV.

Optical studies of single LQDMs confirm that electron states in the $E_0$ shell are typically isolated to individual QDs.\cite{Hermannstadter2010} However, there has been no experimental evidence for the existence of delocalized electron states in higher energy shells.  Our measured PL spectra of both ensembles and single LQDMs establishes the existence of delocalized molecular-like states in excited electron energy shells and qualitatively validates the energy shell model of Peng et al.

\begin{figure}[htb]
\begin{center}
\includegraphics[width=8.0cm]{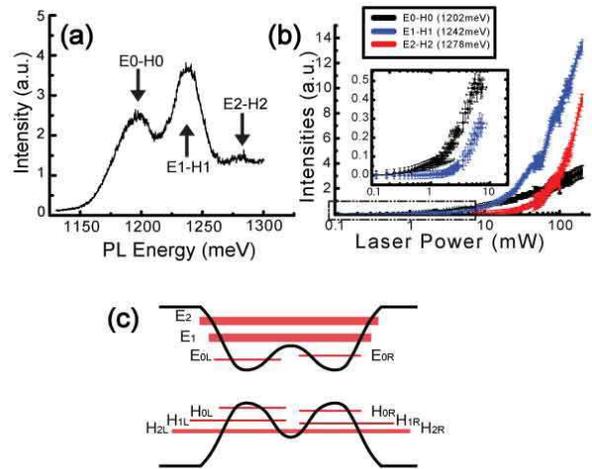}
\caption{(Color Online) a) PL spectra of an ensemble of LQDMs taken in a 5 $\mu$m gap with laser power 30mW and no applied electric field. b) Dependence of PL peak intensity on laser power between 0.1 to 200mW. The inset is the enlarged graph of the area framed with the dashed line rectangle. c) Schematic model of energy level shells as described in text. \label{ensemble}} \end{center}
\end{figure}

In Fig.~\ref{ensemble}a we plot the ensemble PL signal measured in the 5$\mu$m gaps of the interdigitated top contacts. Three PL peaks are visible, centered at 1202, 1242, and 1279 meV. These peaks could be attributed to different energy shells of the LQDMs (analogous to the s, p, and d shells in single InAs QDs) or to a distribution of QD types (e.g. a fraction of the seed QD population that did not evolve into LQDMs). To distinguish these possible explanations, we measure the PL intensity of the three ensemble peaks as a function of laser power. The results are presented in Fig.~\ref{ensemble}b. The inset of Fig.~\ref{ensemble}b shows that the lowest energy peak (1202 meV) dominates emission at the lowest laser power. For laser powers between 0.1 and 2 mW, the intensity of the lowest energy peak increases with laser power and there is no measurable intensity for other peaks. The intensity of the next highest peak (1242 meV) becomes measurable at laser powers of 2 mW. At laser powers greater than 2 mW, the intensity of the peak centered at 1242 meV increases much more rapidly with laser power than the peak centered on 1202 meV. Emission from the highest energy peak (1279 meV) is not measurable before the laser power reaches about 10 mW. Between 10 and 200 mW, the second and third peaks increase in intensity with a similar dependence on laser power.

If the observed PL peaks originated in a distribution of QD types, optical emission from all peaks would be expected at the lowest laser powers. The observation that the lowest energy peak dominates at low powers, and that peaks with increasing energy gain optical intensity only at increasing laser powers, indicates that the three peaks are part of a single structure in which carriers can relax to lower energy shells. Emission from excited states becomes possible only as the increasing laser power generates sufficient carriers to populate the lower energy states and inhibit relaxation. We therefore assign the three observed peaks to different energy shells of single LQDMs. In Fig.~\ref{ensemble}c we schematically depict the LQDM energy shells. The energy levels of the two QDs that comprise the LQDM are typically slightly different; Fig.~\ref{ensemble} depicts a situation in which the left QD has lower energy levels. We stress that the measured ensemble PL presented in Fig.~\ref{ensemble}a and b provides evidence that the three observed PL peaks come from distinct energy shells of a single quantum structure. These data, however, are not sufficient to identify the specific energy shells associated with each transition or the degree of delocalization of states in any shell. As described below, we use additional measurements of single and ensemble LQDMs to systematically justify the model presented in Fig.~\ref{ensemble}c.

We first show that the power dependence and saturation behavior presented in Fig.~\ref{ensemble}b is consistent with assigning the three PL peaks to the $E_0$-$H_0$, $E_1$-$H_1$ and $E_2$-$H_2$ energy shells of LQDMs. At very low exciting laser powers, on average less than one electron and hole are absorbed into the LQDM. These carriers thermally relax through the ladder of energy states in the LQDM. Thermal relaxation in confined structures typically takes place on ps time scales, so emission from excited energy levels is unlikely.\cite{sosnowski1998} Consequently, we assign the lowest energy observed PL peak (1202 meV) to recombination of electrons and holes in their lowest energy states ($E_0$ and $H_0$). The second measured PL peak (1242 meV) could be assigned to optical recombination from $E_0$ to $H_1$, $E_1$ to $H_0$, or $E_1$ to $H_1$. Data presented below demonstrates that the center wavelength of this second PL peak shifts as the charge occupancy of the $E_1$ energy levels changes, so recombination from the $E_0$ shell can be excluded. In single QDs, the asymmetry of envelope functions from different shells (e.g. s and p) suppress optical recombination and observed PL is dominated by s to s and p to p transitions. This argument suggests assigning the peak to recombination from $E_1$ to $H_1$.\footnote{The dipole selection rules are known to be relaxed by strain, piezoelectric effects, and applied electric fields.\cite{Stier1999}. The complex potential profiles of the LQDMs studied here make it likely that the wavefunction symmetries and dipole selection rules are more complicated than those of conventional single QD confined states. Consequently, we cannot rule out recombination from $E_1$ to $H_0$.} The intensity of the $E_1$-$H_1$ peak rises as a function of laser power because increased laser intensity increases the probability that the $E_0$ states will be populated by optically injected carriers. PL emission originating in the $E_1$ energy shell becomes significant only when the $E_0$ states are already filled and relaxation from $E_1$ to $E_0$ is suppressed. By the same argument we assign the third PL peak, centered at 1279 meV, to PL originating in the $E_2$ energy shell.

The measured difference between ensemble PL peak energies $E_0$-$H_0$ and $E_1$-$H_1$ is 40 meV. This value is in reasonable agreement with the separation between $E_0$-$H_0$ and $E_1$-$H_1$ transitions predicted by pseudopotential calculations of LQDMs (approximately 28 meV).\cite{Peng2010a} Similarly, the measured difference between $E_0$-$H_0$ and $E_2$-$H_2$ ensemble PL peak energies (77 meV) is in reasonable agreement with the predicted difference between $E_0$-$H_0$ and $E_2$-$H_2$ transitions (approximately 57 meV ).\cite{Peng2010a} The reasonable agreement with calculated values supports, but does not unambiguously confirm, our assignment of PL peaks. Our measured values of the energy separation between transitions are consistently larger than the those predicted by Peng et al. This difference may be due to differences between the the LQDM morphology used in the computational model and the LQDMs studied experimentally.

\section{Many-body interactions in the $E_1$ energy shell}\label{manybody}

\begin{figure}[htb]
\begin{center}
\includegraphics[width=8.0cm]{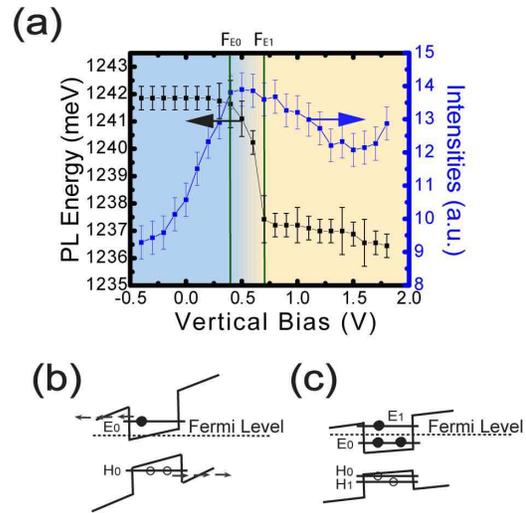}
\caption{(Color Online) (a) PL spectra of the $E_1$-$H_1$ transition as a function of applied voltage. b,c) Schematic models of confining potential along the growth direction with no (b) and significant (c) applied voltage, indicating how the applied voltage impacts the filling of excited energy levels.\label{ens_vs_voltage}} \end{center}
\end{figure}

To understand  the $E_1$ shell, we study the ensemble PL of the $E_1$-$H_1$ transition as a function of applied voltage. Fig.~\ref{ens_vs_voltage}a plots the intensity and center energy of the PL peak as a function of the voltage applied along the growth direction. Because of the Schottky diode structure, zero applied voltage corresponds to a built-in field 19.2 kV/cm. Increasing applied voltage forward biases the diode, dropping the confined energy levels of the QDMs toward the Fermi level set by the doping. As the applied voltage increases, the intensity of PL emission from the $E_1$ state increases until saturating at an applied voltage of about 0.4 V, which we label $F_{E0}$. The center energy of the $E_1$-$H_1$ PL peak remains constant at 1242 meV until $F_{E0}$, then decreases rapidly to about 1237 meV as the applied field is increased to about 0.7 V, which we label $F_{E_1}$. The center PL energy then remains constant at approximately 1237 meV as the applied voltage continues to increase.

To explain the intensity dependence of the $E_1$-$H_1$ PL, we consider the dynamics of electron and hole absorption and relaxation into the $E_1$ energy shells. The number of electrons and holes captured in the LQDM depends on the applied voltage because it impacts the probability with which optically generated charges are captured and retained in the LQDM, as schematically depicted in Fig.~\ref{ens_vs_voltage}b and c. We stress that the band diagrams in Fig.~\ref{ens_vs_voltage}b and c are drawn along the growth axis and thus depict only the vertical confinement in one of the QDs comprising the LQDM. When no voltage is applied, the large built-in electric field creates a relatively thin triangular tunnel barrier for confined electrons and holes, reducing the probability that carriers are trapped in and remain in the QDs (Fig.~\ref{ens_vs_voltage}b). The confinement for holes is stronger than for electrons because of the large hole effective mass. As a result, single InGaAs QDs and vertical QDMs studied under large built-in electric fields frequently exhibit multiple discrete spectral lines associated with positively charged exciton configurations.\cite{Ediger2007, Stinaff2006, Doty2006, Doty2008} The confinement of both electrons and holes is substantially weaker for excited energy states, making it improbable that a carrier population can be trapped in the $E_1$ states.

As the applied voltage is increased, the net electric field decreases and the confining potential becomes more like a square well (Fig.~\ref{ens_vs_voltage}c). Trapping and retention of electrons becomes more probable and the intensity of emission from the $E_1$-$H_1$ PL transition increases. The intensity saturates near the applied voltage at which the $E_0$ states cross the Fermi level ($F_{E0}$) because the $E_0$ states become electrically charged by electrons tunneling from the Fermi level. As a result, at applied voltages larger than $F_{E0}$ all optically generated electrons are populating the $E_1$ states and the probability of emission from this state no longer depends strongly on the applied voltage.

The red shift of the center of the $E_1$-$H_1$ PL peak between $F_{E0}$ and $F_{E_1}$ is explained by charging and many-body interactions within the $E_1$ energy shell. The dynamics of charging described above indicate that, for applied voltages less than $F_{E0}$, the probability that electrons populate $E_1$ is relatively small. As a result, on average only one electron occupies the $E_1$ energy shell. As the applied voltage is increased beyond $F_{E0}$, the probability of electron capture in $E_1$ increases significantly and the average electron occupancy of the $E_1$ energy shell increases. Many-body interactions between the electrons filling the $E_1$ energy shell lead to the red shift in emission of the $E_1$-$H_1$ PL. When the $E_1$ energy shell crosses the Fermi level, the $E_1$ states become fully occupied with electrons tunneling from the Fermi level and the red shift ceases. The ensemble average red shifts by 5 meV between $F_{E0}$ and $F_{E_1}$. The red shift with increasing electron occupancy in consistent with the typical behavior of single InGaAs QDs and VQDMs.\cite{Doty2006a, Ediger2007a}

\section{Spectral signatures of Coulomb interactions}
\label{single}

\begin{figure}[htb]
\begin{center}
\includegraphics[width=6.0cm]{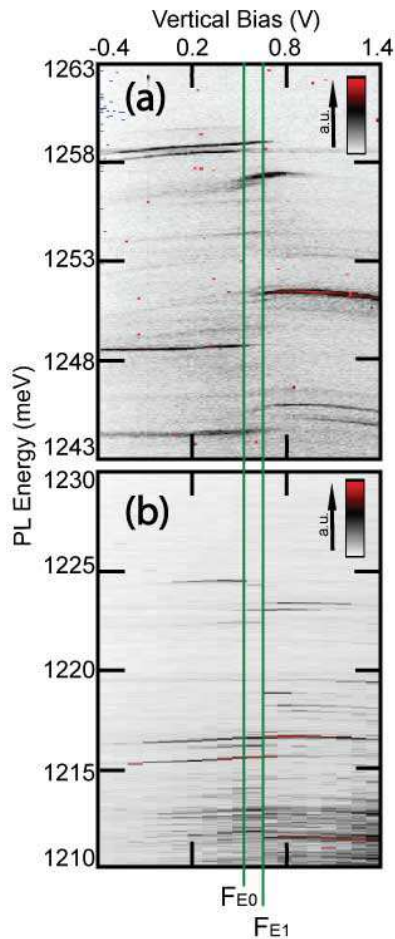}
\caption{(Color Online) PL spectra of the $E_0$-$H_0$ (b) and $E_1$ to $H_1$ (a) transitions in a single LQDM as a function of applied voltage. Vertical lines indicate the applied voltage at which the $E_0$ and $E_1$ energy shells cross the Fermi level. Panel a is measured with about twice of exciting laser power than panel b in order to populate the $E_1$ energy shell. Scale bars in the top right corner indicate the colors associated with increasing PL intensity.\label{E0E1LQDM}} \end{center}
\end{figure}

Further information on the energetic position and degree of localization of energy shells can be obtained from spectroscopy of single LQDMs. We now show that the single LQDM spectra validate the model that electrons in excited energy shells are delocalized while the excited-shell holes remain localized in a single LQD. In Fig.~\ref{E0E1LQDM} we present the discrete spectral lines of a single LQDM measured in both the $E_0$ (b) and $E_1$ (a) energy shells. Vertical lines labeled $F_{E0}$ and $F_{E1}$ indicate applied voltages where discrete shifts in the PL spectra are observed in both energy shells. We attribute the first discrete shift ($F_{E0}$) to Coulomb interactions that arise when the applied voltage tunes the $E_0$ shell below the Fermi level and additional electrons begin to occupy the $E_0$ shell. We attribute the second discrete shift ($F_{E1}$) to charging of the $E_1$ shell. The observation that discrete shifts in the energy of PL transitions within the $E_0$ ($E_1$) energy shell occur when the $E_1$ ($E_0$) energy shell crosses the Fermi level confirms that both of these energy shells are part of the same LQDM. The applied voltages at which the $E_0$ and $E_1$ shells of the single LQDM are charged ($F_{E0} - F_{E1} = 1.2$ V in Fig.~\ref{E0E1LQDM}) correspond to a 6 to 26 meV  separation of the $E_0$ and $E_1$ energy shells of this LQDM. Obtaining a more precise value for the $E_0$-$E_1$ energy shell separation in this LQDM requires knowing the spatial location of the LQDM between the top electrodes, which we cannot measure.

\begin{figure}[htb]
\begin{center}
\includegraphics[width=8.0cm]{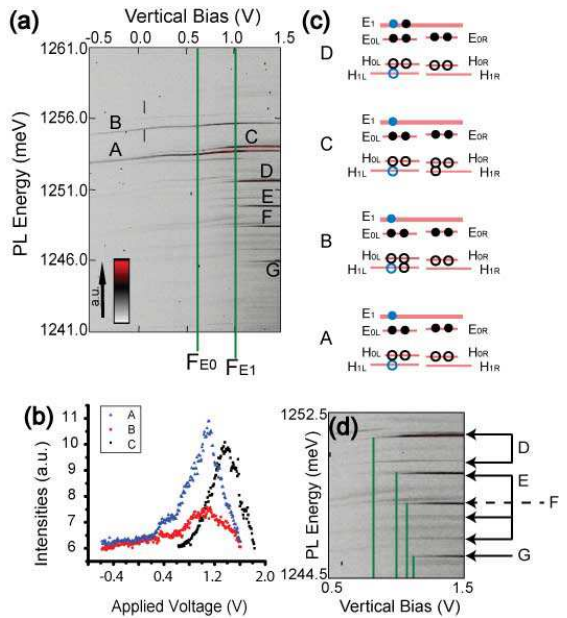}
\caption{(Color Online) a) PL spectrum of the $E_1$-$H_1$ transitions in a single LQDM as a function of applied voltage. The scale bar at the bottom left corner indicates the color mapping of increasing PL intensity. b) Intensity of PL lines A, B and C as a function of applied voltage. c) Depiction of the charge configuration assigned to each labeled transition as described in the text. d) Close up of panel a showing possible spin fine structure correlations. \label{E1H1Single}} \end{center}
\end{figure}

To develop a more detailed understanding of Coulomb interactions and assess the spatial extent of wavefunctions associated with the $E_1$ energy shell, we systematically analyze the spectral map presented in Fig.~\ref{E1H1Single}a. Fig.~\ref{E1H1Single}a plots the observed $E_1$-$H_1$ PL emission of a single LQDM, different from that presented in Fig.~\ref{E0E1LQDM}, as a function of applied voltage. At low values of the applied voltage two discrete PL lines separated by 2.1 meV are observed (A and B). The PL intensity of lines A and B is weak because the probability that electrons occupy the $E_1$ energy level is low. There are three possible origins for this pair of spectral lines: 1) recombination involving energy levels of the two distinct QDs that comprise the LQDM, 2) recombination of two different charge states of the lowest energy QD, and 3) recombination from the lowest energy QD with and without the presence of an extra spectator charge in the other QD.

The A-B peak separation of 2.1 meV is at least four times larger than the spectral shifts typically observed when spectator charges occupy the neighboring QD of a VQDM, where shifts are typically less than 0.5 meV for QDs separated by less than 10 nm. The QD separation in these LQDMs is much larger ($\sim$ 50 nm center-to-center) and the Coulomb interactions between charges localized in separate QDs are therefore expected to be much smaller. As a result, the assignment of the discrete spectral shifts to recombination in the lowest energy QD with and without spectator charges in the neighboring QD can be excluded. Since line A and B show identical shift features with applied voltage, the possibility that these two lines are emitted by two QDMs with different locations can be excluded. We assign the two discrete lines (A and B) to recombination of an electron in the lowest energy $E_1$ state under two different configurations of holes. We analyze the dynamics of charge relaxation into the QDs, the dependence of the relative intensity of PL lines A and B on applied voltage, and the observed spectral shifts to show that the $E_1$ electron state is delocalized while hole states in both the $H_0$ and $H_1$ shells are localized in one of the two QDs that comprise the LQDM (e.g. $H_{1L}$ and $H_{1R}$).

Experiments and calculations on LQDMs indicate that electrons relax more rapidly than holes. Electrons can relax into either of the two QDs that comprise the LQDM (e.g. $E_{0L}$ or $E_{0R}$) and typically relax into the QD with the lowest energy levels.\cite{Hermannstadter2010} The QD with lowest energy levels depends on the applied lateral electric field, but in the conditions studied here the applied lateral field is small and changes only slightly with applied voltage, as discussed above.  We therefore conclude that the difference in energy levels is dominated by the natural variation in QD energy levels arising from growth inhomogeneity. We further assume that the energy ordering of the states associated with the two QDs does not change with applied voltage. We cannot determine which QD has lower energy states, but for consistency in the discussion we will assume that the left QD has lower energy levels, following our schematic in Fig.~\ref{ensemble}c.

Holes relax more slowly than electrons and typically fall into the same QD as the electron due to Coulomb interactions.\cite{Hermannstadter2010} When the electric field is large, optical charging tends to populate positively charged exciton states, as described above. This optical charging is a quasi-random process, and multiple charge states can be observed in the time-integrated PL spectra.

The charge state assignments for transitions labeled in Fig.~\ref{E1H1Single}a are depicted in Fig.~\ref{E1H1Single}c. The electron and hole participating in the excitonic recombination are colored blue and all spectator charges are colored black. We stress that we cannot verify the total charge configuration for each transition. In particular, there may be fewer spectator charges than we depict in Fig.~\ref{E1H1Single}c. However, the \emph{changes} in charge occupancy depicted in Fig.~\ref{E1H1Single}c explain the energy differences between PL lines measured in the data. We first present the charge configurations tentatively assigned to each PL line and show how changing Coulomb interactions in the different configurations predict energy shifts consistent with the measured data. We then show that these measured Coulomb interactions support our conclusion that the $E_1$ energy shell is delocalized while the $H_0$ and $H_1$ shells are localized to single QDs.

In single InGaAs QDs and VQDMs, the addition of a single hole to a single QD typically increases the energy of PL emission from that QD by 2-3 meV.\cite{Stinaff2006, Doty2006, Ediger2007a} We therefore assign PL line A to recombination of an electron from $E_1$ to $H_{1L}$ when the $E_0$ and $H_0$ states are filled and there are no additional holes in $H_{1L}$. We assign PL line B to the same charge configuration as transition A with the addition of one hole in $H_{1L}$. This assignment is supported by the relative intensities of PL lines A and B as a function of applied voltage, as depicted in Fig.~\ref{E1H1Single}b. As the applied voltage increases, the probability of populating the $E_1$ and $H_1$ levels increases, so the intensity of both PL lines increases. However, the probability of generating an excess hole population \emph{decreases} with increasing applied voltage. As a result, the intensity of the positively charged exciton (PL line B) decreases with respect to PL line A.

As the applied voltage in Fig.~\ref{E1H1Single}a increases beyond $F_{E0}$, a new spectral line (C) appears. This line starts with low intensity, but increases in intensity as the applied voltage is increased. The small energy shift between lines A and C ($\sim$ 0.31 meV ) suggests that PL line C arises from Coulomb interactions with a charge localized in the right QD. We attribute this PL line to emission of the $E_1$ to $H_{1L}$ transition in the presence of an additional spectator hole in the right QD (see Fig.~\ref{E1H1Single}c). At applied voltages less than $F_{E0}$ the occupation of the high energy (R) QD by holes is less probable because the electron localizes to the low energy (L) QD and the hole follows. Charging of the $E_{0R}$ and $H_{0R}$ states therefore occurs only when optical excitation injects multiple electron hole pairs. At applied voltages greater than $F_{E0}$, however, electrons occupy both the $E_{0L}$ and $E_{0R}$ states. Consequently, Coulomb interactions no longer favor relaxation of the hole into the low energy (L) QD. The probability of observing a spectator hole in the right QD therefore increases after the applied voltage crosses the Fermi level, consistent with the assignment of PL line C to the $E_1$ to $H_{1L}$ transition in the presence of an additional spectator hole in the right QD.

As the applied voltage in Fig.~\ref{E1H1Single}a increases beyond $F_{E1}$, a series of strong, discrete PL lines appears (D-G). Lines D through G do not all gain significant optical intensity at the same value of applied voltage. Rather, line D become optically bright at 1.00 V, with lines E-G gaining optical intensity at 1.08, 1.17, 1.31 V respectively. This sequential appearance of the PL lines suggests that the lines are associated with the sequential charging of the $E_1$ states with single electrons as the states cross the Fermi level. The charge configuration for PL line D is depicted in Fig.~\ref{E1H1Single}c and lines E, F, and G are assigned to filling of the $E_1$ energy levels with 3, 4, and 5 electrons, respectively. The applied voltage at which each additional electron enters the LQDM depends on both the exact energy level of each state within the $E_1$ shell and Coulomb interactions with the other electrons occupying the LQDM. Each time a charge is added, the PL energy red shifts by approximately 2 meV (A-D: 2.25 meV, D-E: 1.87 meV, E-F: 1.53 meV, F-G: 2.59 meV. The red shift is consistent with the energy shift observed in other InGaAs QDs upon charging with additional electrons.\cite{Ediger2007a}

If two charges are localized in separate QDs, the wavefunctions associated with the charges have minimal overlap. Consequently, the shift in PL energy due to Coulomb interactions between charges localized in separate QDs is small, of order 0.5 meV. Shifts of this magnitude are observed for PL line C in Fig.~\ref{E1H1Single}, supporting the conclusion that the hole states are localized in individual QDs. However, all red shifts observed upon charging with additional electrons (lines D-G) are of order 2 meV. This large shift is indicative of strong Coulomb interactions and suggests that the electron wavefunctions overlap strongly. We thererfore conclude that the $E_1$ states are delocalized over the entire LQDM. If states of the $E_1$ shell are delocalized, states of the $E_2$ shell must also be delocalized. This analysis is the basis of the schematic depiction of delocalized states in Fig.~\ref{ensemble}c. We note that the localization or delocalization of the $H_2$ states cannot be determined from the presently available data.

\section{Spin fine structure in a single LQDM}
\label{finestructure}

Further evidence for the delocalized nature of the $E_1$ energy levels is provided by the observation of fine structure in the highly charged exciton states (D-G), as shown in Fig.~\ref{E1H1Single}d. Fine structure typically arises in PL spectra when more than one spin configuration is possible in either the initial or final states. Electrons have spin projections $\pm \frac{1}{2}$. Holes in QDs are properly described as spinors with both light and heavy hole character, though the heavy hole component typically dominates the spinor.\cite{Climente2008} The light hole components can have a significant impact on the properties of coupled QDs,\cite{Doty2010a, Planelles2010} but the assumption that holes have only heavy hole character with spin projections $\pm \frac{3}{2}$ is a reasonable first approximation. Optical selection rules require recombining electrons and holes to have opposite spin projections, so dark exciton states, where the electron and hole have parallel spin projections, are not observed.\footnote{We ignore anisotropic exchange interactions, which typically introduce energy splittings of order 100 $\mu$eV, comparable to the resolution of our spectrometer.} The Pauli exclusion principle requires two charges occupying a single energy level to be in a spin singlet configuration. As a result, neutral exciton and singly charged (negative or positive) exciton transitions do not have spin fine structure. Spin fine structure becomes possible only in charged exciton states where there are two unpaired spins in either the initial or final state of optical recombination.

Transitions A-C in Fig.~\ref{E1H1Single} all involve recombination where spin fine structure is not expected. Neither the initial nor final state of transition B has more than one unpaired spin. The initial state of transition A has an unpaired electron and unpaired hole, but the dark exciton spin configurations cannot be observed. Transition C has two unpaired holes, but experiments on VQDMs indicate that singlet and triplet configurations of holes localized in separate QDs are degenerate in energy unless coherent tunneling between the energy levels leads to a kinetic exchange interaction.\cite{Scheibner2007} Because the two QDs that comprise the LQDM typically have slightly different energy levels, coherent coupling is unlikely in the absence of any lateral field to tune the energy levels into resonance.\cite{Hermannstadter2010} The absence of spin fine structure associated with line C further supports the conclusion that the $H_1$ energy levels are localized to individual QDs.

Once the applied voltage tunes the $E_1$ shell past the Fermi level, electrons begin to fill the states that comprise the $E_1$ shell. The order in which these states will be filled depends on charge and spin interactions.\cite{Ediger2007a} Whenever the initial state or final state of optical recombination has unpaired electron spins, multiple possible spin configurations are possible and spin fine structure is probable. In Fig.~\ref{E1H1Single}d we indicate several weak PL lines that gain optical intensity at the same value of applied voltage as the strong PL lines D and E. Although we cannot assign these spectral lines to specific charge and spin configurations, the applied voltage controls the number of electrons that occupy the $E_1$ shell. Consequently, a pair of lines that gain optical intensity at the same value of applied voltage most probably originate from states with the same total charge. The observation of fine structure for line D suggests that the two electrons occupying the $E_1$ shell are located in different discrete energy states. If these states were localized in individual QDs, the energy shifts due to Coulomb or spin interactions would be expected to be very small (0.5 meV or less). The observed shifts are all more than 1meV (D: 1.27 meV, E: 2.16 and 3.23 meV), further supporting the conclusion that the $E_1$ energy states are delocalized over the entire LQDM.

Unlike controlled charging of single InGaAS QDs,\cite{Ediger2007a} there is not a well-defined window of applied voltage in which the PL of a single charge configuration dominates. This suggests that the charge occupancy of the LQDM $E_1$ state is not well defined for any value of applied field. States with increasing charge occupancy do not become populated until the applied voltage tunes the $E_1$ energy states sufficiently below the Fermi level to overcome Coulomb interactions and add an additional electron to the $E_1$ energy shell. For this reason the PL transitions originating in states with increasing number of electrons gain optical intensity at increasing values of the applied voltage. However, optical recombination of electrons in the $E_1$ states competes against emission from the $E_0$ states, which removes electrons and enables electron relaxation out of the $E_1$ states. Consequently, states of lower charge occupancy can continue to emit even at applied voltages where higher charge occupancy is possible.

\section{Conclusion}\label{conclusion}
Spectroscopy of an ensemble of LQDMs reveals three PL peaks centered on 1202 meV, 1242 meV and 1279 meV. By analyzing the intensity and center energy of these peaks as a function of laser power and applied voltage, we attribute these peaks to optical recombination of electrons and holes in different energy shells of the LQDMs. Spectral maps of single LQDMs as a function of applied voltage confirm this assignment. The spectroscopy of single LQDMs also allows us to identify the characteristic spectral shifts associated with electron charging and Coulomb interactions. The magnitude of these shifts supports the conclusion that the $E_0$, $H_0$ and $H_1$ energy shells are localized to individual QDs while the $E_1$ energy shell is delocalized over the entire LQDM. The observation of fine structure that may be associated with spin interactions further validates this model. The delocalized $E_1$ electron states observed here may provide a mechanism for mediating coherent interactions between spins localized in the $E_0$ or $H_0$ energy levels of the two distinct QDs that comprise the LQDM.

\section{Acknowledgement}\label{acknowledgement}
 The authors would like to acknowledge Hongtao Lin and Juejun Hu for simulations of the electrical potential of the three-electrodes device. The work was financially supported by NSF DMR-0844747 for Zhou/Doty and Basic Science Program through the National Research Foundation (NRF) of Korea funded by the Ministry of Education, Science and Technology (Grant No. 2011-0004804) for Lee.


\end{document}